\documentclass[aps,prl,twocolumn,superscriptaddress,showpacs]{revtex4}
\usepackage{amssymb}
\usepackage{txfonts}
\usepackage{tipa}

\usepackage{graphicx}
\usepackage{dcolumn}
\usepackage{bm}

\begin{document}


\title{Magnetic structure and orbital ordering in KCrF$_3$ perovskite}

\author{Y. Xiao}
\email[Electronic address: y.xiao@fz-juelich.de]{}
\affiliation{Institut fuer Festkoerperforschung, Forschungszentrum
Juelich, D-52425 Juelich, Germany}

\author{Y. Su}
\affiliation{Juelich Centre for Neutron Science, IFF,
Forschungszentrum Juelich, Outstation at FRM II, Lichtenbergstr. 1,
D-85747 Garching, Germany}

\author{H.-F. Li}
\affiliation{Institut fuer Festkoerperforschung, Forschungszentrum
Juelich, D-52425 Juelich, Germany} \affiliation{Ames Laboratory and
Department of Physics and Astronomy, Iowa State University, Ames,
Iowa 50011, USA}

\author{C.M.N. Kumar}
\affiliation{Institut fuer Festkoerperforschung, Forschungszentrum
Juelich, D-52425 Juelich, Germany}

\author{R. Mittal}
\affiliation{Juelich Centre for Neutron Science, IFF,
Forschungszentrum Juelich, Outstation at FRM II, Lichtenbergstr. 1,
D-85747 Garching, Germany}

\affiliation{Solid State Physics Division, Bhabha Atomic Research
Centre, Trombay, Mumbai 400085, India}

\author{J. Persson}
\affiliation{Institut fuer Festkoerperforschung, Forschungszentrum
Juelich, D-52425 Juelich, Germany}

\author{A. Senyshyn}
\affiliation{Institute for Materials Science, Darmstadt University
of Technology, D-64287 Darmstadt, Germany}
\affiliation{Forschungsneutronenquelle Heinz Maier-Leibnitz (FRM
II), D-85747 Garching, Germany}

\author{K. Gross}
\affiliation{Institut fuer Festkoerperforschung, Forschungszentrum
Juelich, D-52425 Juelich, Germany}

\author{Th. Brueckel}
\affiliation{Institut fuer Festkoerperforschung, Forschungszentrum
Juelich, D-52425 Juelich, Germany} \affiliation{Juelich Centre for
Neutron Science, IFF, Forschungszentrum Juelich, Outstation at FRM
II, Lichtenbergstr. 1, D-85747 Garching, Germany}

\date{\today}

\begin{abstract}
KCrF$_3$ represents another prototypical orbital-ordered perovskite,
where Cr$^{2+}$ possesses the same electronic configuration of
3\emph{d}$^4$ as that of strongly Jahn-Teller distorted Mn$^{3+}$ in
many CMR manganites. The crystal and magnetic structures of KCrF$_3$
compound are investigated by using polarized and unpolarized neutron
powder diffraction methods. The results show that the KCrF$_3$
compound crystallizes in tetragonal structure at room temperature
and undergoes a monoclinic distortion with the decrease in
temperature. The distortion of the crystal structure indicates the
presence of cooperative Jahn-Teller distortion which is driven by
orbital ordering. With decreasing temperature, four magnetic phase
transitions are observed at 79.5, 45.8, 9.5, and 3.2 K, which
suggests a rich magnetic phase diagram. Below \emph{T}$_N$ = 79.5 K,
the Cr$^{2+}$ moment orders in an incommensurate antiferromagnetic
arrangement, which can be defined by the magnetic propagation vector
($\frac{1}{2}$$\pm\,$$\delta\,$, $\frac{1}{2}$$\pm\,$$\delta\,$, 0).
The incommensurate-commensurate magnetic transition occurs at 45.8 K
and the magnetic propagation vector locks into ($\frac{1}{2}$,
$\frac{1}{2}$, 0) with the Cr moment of 3.34(5) $\mu_B \,$ aligned
ferromagnetically in (220) plane, but antiferromagnetically along
[110] direction. Below 9.5 K, the canted antiferromagnetic ordering
and weak ferromagnetism arise from the collinear antiferromagnetic
structure, while the Dzyaloshinskii-Moriya interaction and tilted
character of the single-ion anisotropy might give rise to the
complex magnetic behaviors below 9.5 K.

\end{abstract}

\pacs{75.25.Dk, 75.25.-j, 75.50.Ee, 71.70.Ej}
\maketitle

The strongly correlated electron systems such as 3\emph{d}
transition metal oxides have attracted considerable attention due to
their complex structural, electronic and magnetic characteristics
\cite{Imada, Lee}. The interactions between lattice, spin and
orbital degrees of freedom are considered to play the essential role
in explaining some interesting physical properties \cite{Tokura1}.
For example, the colossal magnetoresistance (CMR) effect was
observed in hole-doped perovskite manganites such as
La$_{1-x}$\emph{M}$_x$MnO$_3$(\emph{M} = Sr, Ca and Ba)
\cite{Tokura2, Asamitsu, Jin, Helmolt}. While the undoped parent
compound LaMnO$_3$ is known to be an \emph{A}-type antiferromagnetic
(AFM) insulator in which the orbital ordering is formed due to the
cooperative Jahn-Teller (JT) effect. The electronic configuration of
Mn$^{3+}$ ions in LaMnO$_3$ is $t_{2g}^3$$e_{g}^1$ by using the
Hund's rule as a first approximation. The three electrons in the
$t_{2g}$ orbitals are localized with a total spin 3/2 while the
electron in the $e_{g}$ atomic orbitals is strongly hybridized with
the neighboring oxygen \emph{p} obitals. This particular orbital
ordering is responsible for the \emph{A}-type antiferromagnetic
structure of LaMnO$_3$ \cite{Wollan}. The mixed valence
(Mn$^{3+}$Mn$^{4+}$) will be formed with the hole doping on the
\emph{e}$_g$ band. It is believed that the CMR effect is due to the
double exchange of electrons between ferromagnetically coupled
Mn$^{3+}$ and Mn$^{4+}$ ions \cite{Tokura2}. Therefore, the
interplay and coupling between the lattice, orbital and spin degrees
of freedom are interesting and deserve to be investigated from both
the fundamental and technical point of view.

Besides the transition metal oxides, the transition metal fluorides
also exhibit the intriguing electronic and magnetic effects.
However, compared with the extensive study of oxides, the study of
fluorides is still lack due to the difficulty of synthesis. KCrF$_3$
is one kind of perovskite structure fluorides in which the
electronic and structural characteristics is expected to resemble
with those of widely studied LaMnO$_3$ compounds since the orbital
degrees of freedom are activated for Cr$^{2+}$ (\emph{d}$^4$) cation.
Recently, the crystal structure and magnetic properties of KCrF$_3$
compound were investigated by synchrotron x-ray powder diffraction
and magnetic measurements \cite{Margadonna1}. Both the structural and
magnetic phase transitions seem to be more complex than expected.
The KCrF$_3$ displays not only the large cooperative Jahn-Teller
distortions at room temperature but also a series of temperature
induced structural and magnetic transformations. However, the
detailed magnetic structure of KCrF$_3$ is still not studied
systematically. To clarify the magnetic ordering is of important
significance to understand the interactions between the lattice,
spin and orbital degrees of freedom in KCrF$_3$.

In this paper, we have studied both the cooperative Jahn-Teller
distortion and magnetic phase transition of KCrF$_3$ by using
polarized and unpolarized neutron powder diffraction (NPD) methods.
We have observed series of magnetic phase transitions including an
incommensurate-commensurate magnetic phase transition. The magnetic
phase diagram of KCrF$_3$ is presented based on the magnetization,
heat capacity and NPD measurements between 2.5 and 300 K .

\begin{figure}
\includegraphics[width=8cm,height=11cm]{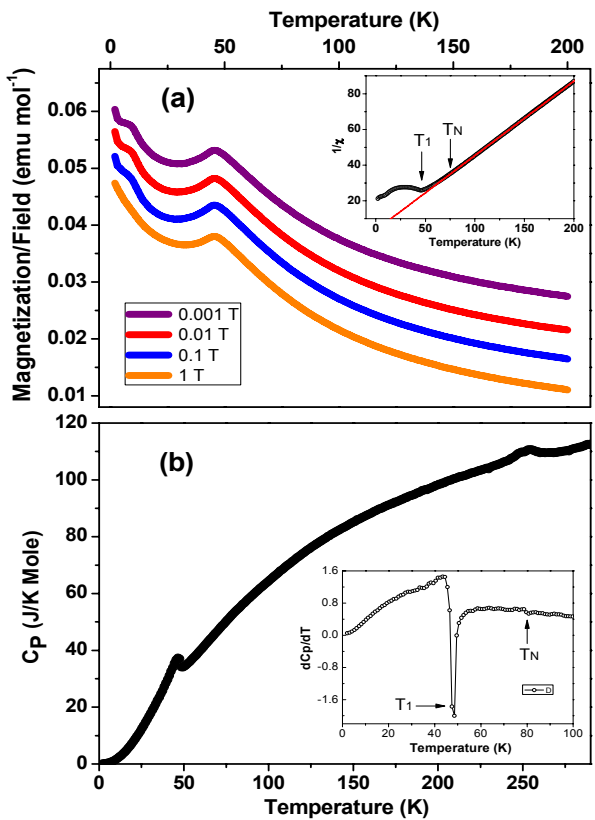}
\caption{\label{fig:epsart} (Color online) (a) Temperature
dependence of the magnetization of KCrF$_3$ measured in
zero-field-cooling mode under different magnetic fields. In order to
clearly show the change of magnetic features, the unit is allocated
only to magnetization curve measured with applied field of 1 T,
while the rest three curves ( with fields equal to 0.001, 0.01 and
0.1 T) are shifted upwards. The inset shows the temperature vs.
inverse susceptibility of KCrF$_3$ measured at magnetic field of 0.1
T. (b) Temperature dependence of heat capacity of KCrF$_3$. Inset:
dC$_P$/dT emphasizes breaks in slope of C$_P$(T) data.}
\end{figure}

Polycrystalline KCrF$_3$ sample was prepared by standard solid state
reaction method. The stoichiometeric powders of KF and CrF$_2$ were
mixed, pressed into pellets and sealed in an evacuated quartz tube,
followed by sintering at 700 K for 24 \emph{h} with one intermediate
grinding. All the weighting, mixing and pressing procedures were
performed in a MBRAUN-MB20G glove box with a protective argon
atmosphere. The laboratory x-ray diffraction measurement reveals
that single phase was obtained with small amount of Cr$_2$O$_3$
impurities ($\sim$ 2\%). Temperature dependence of both
magnetization and specific heat were collected by using a Quantum
Design Magnetic Properties Measurement System (MPMS) and a Physical
Properties Measurement System (MPMS), respectively, with temperature
down to 2 K. Neutron diffraction measurements were performed on the
high resolution diffractometer SPODI and the diffuse neutron
scattering spectrometer DNS at FRMII (Munich, Germany). For the NPD
measurement on SPODI, a Ge(551) monochromator was used to produce a
monochromatic neutron beam of wavelength 1.5481 $\buildrel _\circ
\over {\mathrm{A}}$. The polarized analysis was performed on DNS
with the polarized neutron beam of wavelength 4.74 $\buildrel _\circ
\over {\mathrm{A}}$. The program FULLPROF \cite{Rodriguez1} was used
for the Rietveld refinement of the crystal and the magnetic
structures of the compounds.

The temperature dependence of the magnetization of the KCrF$_3$
sample was measured under different magnetic fields as shown in Fig.
1(a). Three phase transitions are clearly visible at 45.8 K
(\emph{T}$_1$), 9.5 K (\emph{T}$_2$) and 3.2 K (\emph{T}$_3$). The
similar discontinuities in the \emph{M-T} curve are also reported in
previous literatures \cite{Margadonna1, Yoneyama} and the anomaly at
45.8 K was considered as the Neel temperature \emph{T}$_N$ where the
paramagnetic state develops from the antiferromagnetic phase with
increasing temperature. However, in present work the polarized
neutron analysis experiment indicated that the kink at 45.8 K
corresponds to a commensurate-incommensurate antiferromagnetic phase
transition and the system enters the paramagnetic state at a higher
temperature of 79.5 K, as discussed in the following text. It is
also noticed that the anomalies at 9.5 and 3.2 K are not shown by
applying higher magnetic field of 1 T which is the signature of the
field-induced magnetic phase transition. The inset of Fig. 1(a)
shows the temperature dependence of the inverse susceptibility
1/$\chi$ of the compound measured under magnetic field of 0.1 T. The
susceptibility strictly follows the Curie-Weiss behavior above 80 K.
The effective paramagnetic moment and Weiss temperature are deduced
to be 4.379(1) $\mu_B \,$ and -8.7(4) K, respectively. It is noticed
that a positive Weiss temperature ($\sim$2.7 K) is given in
Ref.\cite{Margadonna1} based on the analysis of magnetic
susceptibility $\chi$ which is measured in an applied field of 1 T.
It is known that Curie-Weiss law is only valid with low applied
field and too strong field may affect considerably the
susceptibility by changing the weak electron coupling and leading to
magnetic saturation. Therefore, the different Weiss temperatures
obtained for KCrF$_3$ are attributed to different measuring
conditions and the Weiss temperature deduced with lower field
measurement reflects the antiferromagnetic background magnetism in
KCrF$_3$. Fig. 1(b) shows the temperature dependence of the heat
capacity result in which a clear phase transition is detected at 248
K. The crystal structure analysis given below reveals that the kink
at 248 K is associated with the monoclinic to tetragonal crystal
structural phase transition. Another phase transition at 79.5 K can
be more clearly identified by examining the dC$_P$/dT plot as shown
in the inset of Fig. 1(b). Therefore, a series of phase transitions
in KCrF$_3$ are constructed by magnetic and thermal
characterizations, which suggest a rich phase diagram of KCrF$_3$.

\begin{figure}
\includegraphics[width=8.6cm,height=10.0cm]{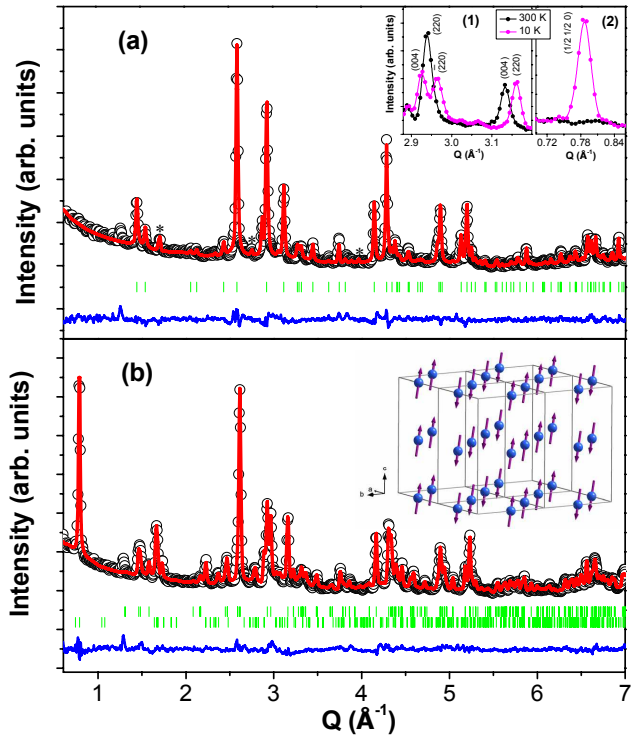}
\caption{\label{fig:epsart} (Color online) (a) NPD refinement
pattern for the KCrF$_3$ at 300 K. The vertical bars at the bottom
indicate the Bragg reflection positions, and the lowest curve is the
difference between the observed and the calculated NPD patterns.
Cr$_2$O$_3$ impurities are marked with asterisks. The Inset (1) and
(2) show the splitting of tetragonal (220) peak and the onset of
($\frac{1}{2}$$\frac{1}{2}$0) magnetic reflection at low
temperature, respectively. (b) NPD refinement pattern for the
KCrF$_3$ at 10 K. Inset shows the magnetic structure of KCrF$_3$ at
10 K. Frames indicate the crystallographic unit cells and the
magnetic unit cell doubled along both \emph{a} and \emph{b}
directions.}
\end{figure}

\begin{table}
\caption{\label{tab:table1} Refined results of the crystal and
magnetic structures for KCrF$_3$ at 10 and 300 K. The atomic
positions for space group \emph{I112/m}: K(4\emph{g})(0,0,\emph{z}),
Cr$_{(1)}$(2\emph{c})(0,0.5,0), Cr$_{(2)}$(2\emph{d})(0,0.5,0.5),
F$_{(1)}$(4\emph{i})(\emph{x},\emph{y},0),
F$_{(2)}$(4\emph{i})(\emph{x},\emph{y},0),
F$_{(3)}$(4\emph{h})(0,0.5,\emph{z}); for \emph{I4/mcm}:
K(4\emph{a})(0,0,0.25), Cr$_{(1)}$(4\emph{d})(0,0.5,0),
F$_{(1)}$(4\emph{b})(0,0.5,0.25),
F$_{(2)}$(8\emph{h})(\emph{x},\emph{y},0).}
\begin{ruledtabular}
\begin{tabular}{lll}
Temperature & 10 K & 300 K\\
\hline

Space group & \emph{I112/m} & \emph{I4/mcm}\\

\emph{a} \, ($\buildrel _\circ \over {\mathrm{A}}$)&5.8069(7)&6.0464(6)\\
\emph{b} \, ($\buildrel _\circ \over {\mathrm{A}}$)&5.8137(7)&6.0464(6)\\
\emph{c} \, ($\buildrel _\circ \over {\mathrm{A}}$)&8.5871(7)&8.0230(8)\\
\emph{$\gamma$} \, ($^\circ$)&93.671(5)&90\\
\emph{V} \, ($\buildrel _\circ \over {\mathrm{A}}$$^3$)&289.303(2)&293.312(2)\\

K \\
\quad \quad \, $\emph{z}$&0.247(2)&0.25\\
\quad \quad \emph{B} ($\buildrel _\circ \over {\mathrm{A}}$$^2$)&1.3(2)&2.4(2)\\

Cr$_{(1)}$ \\
\quad \quad \emph{B} ($\buildrel _\circ \over {\mathrm{A}}$$^2$)&1.1(2)&1.8(1)\\
\quad \quad \emph{M}$$($\mu$$_B$)&3.34(5)&\\

Cr$_{(2)}$ \\
\quad \quad \emph{B} ($\buildrel _\circ \over {\mathrm{A}}$$^2$)&1.4(2)&\\
\quad \quad \emph{M}$$($\mu$$_B$)&3.34(5)&\\

F$_{(1)}$ \\
\quad \quad \, $\emph{x}$&0.291(2)&0\\
\quad \quad \, $\emph{y}$&0.718(2)&0.5\\
\quad \quad \emph{B} ($\buildrel _\circ \over {\mathrm{A}}$$^2$)&1.6(2)&3.0(2)\\

F$_{(2)}$ \\
\quad \quad \, $\emph{x}$&0.230(2)&0.231(2)\\
\quad \quad \, $\emph{y}$&0.197(2)&0.731(2)\\
\quad \quad \emph{B} ($\buildrel _\circ \over {\mathrm{A}}$$^2$)&1.7(2)&2.6(2)\\

F$_{(3)}$ \\
\quad \quad \, $\emph{z}$&0.230(2)&\\
\quad \quad \emph{B} ($\buildrel _\circ \over {\mathrm{A}}$$^2$)&1.7(2)&\\

\emph{R$_p$}&2.76&2.70\\
\emph{R$_{wp}$}&3.56&3.51\\
$\chi$$^2$&3.51&3.35\\

\end{tabular}
\end{ruledtabular}
\end{table}

The neutron powder diffraction pattern of KCrF$_3$ at 300 K is shown
in Fig. 2(a). The compound crystallizes in the tetragonal phase with
space group \emph{I4/mcm}. The lattice parameters are deduced to be
\emph{a} = 6.0464(6) $\buildrel _\circ \over {\mathrm{A}}$ and
\emph{c} = 8.0230(8) $\buildrel _\circ \over {\mathrm{A}}$, which is
in good agreement with that derived from synchrotron x-ray powder
diffraction \cite{Margadonna1}. The detailed structural information
for KCrF$_3$, as obtained from NPD data, are given in Table I. The
crystal structure can be described as the stacking of layers of
corner-sharing CrF$_6$ octahedra. In the \emph{ab} plane the CrF$_6$
octahedra are elongated along the \emph{a} or \emph{b} axis in an
alternative pattern. The structure distortion is caused by the
orbital order associated with the cooperative Jahn-Teller
distortion, \emph{i.e.} the $d_{3x^2-r^2}$/$d_{3y^2-r^2}$ orbitals
are stabilized by the Jahn-Teller distortion and order in an
alternate staggered pattern in the \emph{ab} plane as illustrated in
Fig. 3(a). Similar to the MnO$_6$ octahedral distortion in LaMnO$_3$
\cite{Chatterji, Rodriguez2}, there also exist three Cr-F distances
in KCrF$_3$ named short \emph{L}$_s$, medium \emph{L}$_m$ and long
\emph{L}$_l$. The magnitude of the Jahn-Teller distortion of CrF$_6$
can be evaluated by the following equation: $\Delta$ =
(1/6)$\sum$[(\emph{d}$_n$-$\langle$\emph{d}$\rangle$)/$\langle$\emph{d}$\rangle$]$^2$,
where $\langle$\emph{d}$\rangle$ and \emph{d}$_n$ are the mean Cr-F
bond length and the six Cr-O bond lengths along six different
directions, respectively. The calculated value of $\Delta$ is
4.87$\times$10$^{-3}$ for KCrF$_3$ at 300 K. With decreasing
temperature, the KCrF$_3$ undergoes a monoclinic phase transition as
revealed by the NPD pattern at 10 K. The inset(1) of Fig. 2(a) shows
the obvious splitting of the tetragonal (220) peak into the
monoclinic (220) and ($\bar{2}$20) peaks. Along with the monoclinic
distortion, two inequivalent Cr sites (2\emph{c} and 2\emph{d}) form
with similar octahedral environment. The Jahn-Teller distortion
parameter was deduced to be 5.66$\times$10$^{-3}$  and
3.85$\times$10$^{-3}$ for 2\emph{d} and 2\emph{c} CrF$_6$ octahedra,
respectively. The alternated arrangement of CrF$_6$ octahedra within
the \emph{ab} plane also indicates the change in orbital ordering as
illustrated in Fig. 3(b). However, it should be noted that the
change in magnetic order and the change in orbital order took place
at different temperatures, \emph{i.e.} there is no direct
correlation between orbital ordering and magnetic ordering in
KCrF$_3$. The correlation between magnetic ordering and orbital
ordering are also investigated in LaMnO$_3$ by Sub\'{\i}as \emph{et
al.} \cite{Sub} with resonant x-ray scattering method, which can act
as a direct probe for orbital ordering. Their experimental results
also indicated that there is no correlation between resonant
scattering and long-range AFM ordering in LaMnO$_3$.

\begin{figure}
\includegraphics[width=8cm,height=10cm]{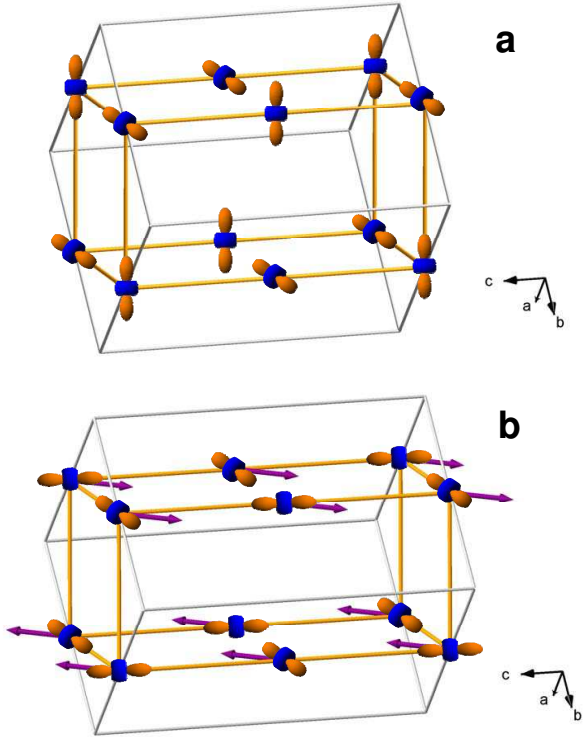}
\caption{\label{fig:epsart} (Color online) Schematic view of the
orbital and spin ordering in the perovskite chromite KCrF$_3$ of
tetragonal (a) and monoclinic (b) phases. The grey line outlines the
unit cell.}
\end{figure}

Besides the crystal structural phase transition, the magnetic phase
transition is also observed at 10 K due to the ordering of the
Cr$^{2+}$ moments as revealed by the onset of magnetic reflection
(inset(2) in Fig. 2(a)). The magnetic reflection at Q = 0.7904
$\buildrel _\circ \over {\mathrm{A}}$$^{-1}$ can be indexed
accurately with ($\frac{1}{2}$$\frac{1}{2}$0) instead of (001)
magnetic reflection \cite{Scatturin} although the \emph{d} values of
those two reflections are quite close. It strongly indicated that
the moments of Cr$^{2+}$ formed antiferromagnetic structure with
($\frac{1}{2}$,$\frac{1}{2}$,0) magnetic modulation. Also, the
existence of considerable intensity of ($\frac{1}{2}$$\frac{1}{2}$2)
magnetic reflection indicates that the Cr$^{2+}$ moment tilts
outward \emph{c} axis. By performing refinement on the NPD data, we
found that the magnetic structure of KCrF$_3$ can be described by
the colinear antiferromagnetic structure model, \emph{i.e.} the
KCrF$_3$ system develops a long-range antiferromagnetic order at 10
K with the Cr$^{2+}$ spins coupled ferromagnetically in the (220)
plane and antiferromagnetically along the [110] direction as shown
in the inset of Fig. 2(b). The moment of Cr$^{2+}$ at 10 K is
deducted to be 3.34(5) $\mu_B \,$ which is close to the theoretical
value \cite{Giovannetti, Xu}. Similar to the \emph{A}-type AFM in
LaMnO$_3$ \cite{Wollan, Solovyev, Feinberg}, the layered AFM
structure of KCrF$_3$ can also be considered as the consequence of
an interplay between superexchange and Jahn-Teller coupling.
According to the Goodenough-Kanamori-Anderson (GKA) rules
\cite{Kanamori, Anderson, Goodenough, Kugel, Khomskii}, the
superexchange between magnetic ions is mediated by the F ligand and
the FM coupling is expected due to the superexchange interaction
between the filled $e_{g}$ and empty orbitals, while the AFM
coupling is expected due to the superexchange interaction between
the unfilled orbitals. Our experimental results confirmed the
picture of the interactions between magnetic ordering and orbital
ordering given by the GKA rules. The schematic view of the orbital
ordering and spin ordering in the unit cell of monoclinic KCrF$_3$
is shown in Fig. 3(b).

\begin{figure}
\includegraphics[width=8.5cm,height=4.5cm]{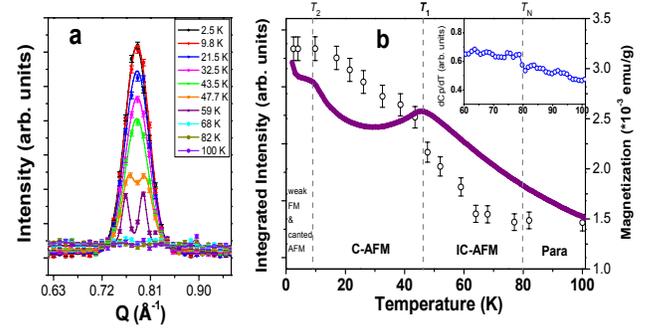}
\caption{\label{fig:epsart} (Color online) (a) Evolution of the
($\frac{1}{2}$$\frac{1}{2}$0) magnetic reflection with the change of
temperature. (b) Magnetic phase diagram of KCrF$_3$. The open circle
and purple curve represent the temperature dependences of integrated
intensity of ($\frac{1}{2}$$\frac{1}{2}$0) magnetic reflection and
magnetization, respectively. The inset shows the enlarged view of
the slope of C$_P$ data.}
\end{figure}

In order to inspect the evolution of the magnetic ordering and
clarify magnetic structure of the KCrF$_3$ system, neutron
polarization analysis was performed at DNS in the \emph{Q} range
from 0.346 to 2.29 $\buildrel _\circ \over {\mathrm{A}}$$^{-1}$ and
temperature range from 2.5 to 100 K with smaller temperature steps.
The nuclear coherent, spin-incoherent and magnetic scattering cross
sections can be separated with the \emph{xyz}-polarization method in
the spin-flip and non-spin-flip channels with polarized
analysis\cite{Schweika}. At 9.8 K, magnetic reflections are clearly
seen and match with the presented AFM structure model. The evolution
of the ($\frac{1}{2}$$\frac{1}{2}$0) magnetic reflection is plotted
in Fig. 4(a) and the integrated intensity of
($\frac{1}{2}$$\frac{1}{2}$0) magnetic reflection in Fig. 4(b) as
the open circles. Here we will discuss the magnetic structure below
10 K firstly. Compared to the ($\frac{1}{2}$$\frac{1}{2}$0)
reflection at 9.8 K, no obvious change in integrated intensity was
detected for the ($\frac{1}{2}$$\frac{1}{2}$0) reflection at 4.0 and
2.5 K, which may suggest a slight canting of the magnetic moment of
Cr$^{2+}$ ions outward the (220) ferromagnetic plane. It is known
that canted AFM and weak FM moment are observed in both
noncrystalline and single crystal LaMnO$_3$ samples \cite{Matsumoto,
Skumryev}. The FM moment in LaMnO$_3$ is caused by the tilt of
Mn$^{3+}$ moments away from the AFM \emph{ab} plane resulting in a
net magnetization along \emph{c} axis \cite{Skumryev, Talbayev}.
Considering the structural comparability between the KCrF$_3$ and
LaMnO$_3$ systems, the increasing magnetization below 9.5 K in
KCrF$_3$ may arise due to the tilt of the Cr$^{2+}$ moment out of
the (220) plane. Both the Dzyaloshinskii-Moriya (DM) exchange
interaction and the character of the single-ion anisotropy are
necessary to give an explanation on that weak ferromagnetism
\cite{Skumryev, Talbayev, Dzyaloshinsky, Moriya}. Therefore, the
magnetic structure below 9.5 K is associated with the mixed state of
weak FM and canted AFM configurations. The neutron diffraction
analysis on KCrF$_3$ single crystal is required to explore the exact
magnetic structure of KCrF$_3$ below 10 K and to illustrate the
nature of phase transition at 3.2 K.

Now we discuss the magnetic structure above 10 K. As shown in Fig.
4(a), the intensity of the ($\frac{1}{2}$$\frac{1}{2}$0) magnetic
reflection decreases gradually with increasing temperature from 9.8
K, which suggests the decrease of amplitude of the Cr$^{2+}$ moment.
At 47.7 K, two satellite reflections are developed instead of the
single ($\frac{1}{2}$$\frac{1}{2}$0) reflection. It suggests that
the transition at 45.8 K corresponds to the magnetic phase
transition from simple collinear AFM phase to an incommensurate
magnetic phase with the propagation vector
($\frac{1}{2}$$\pm\,$$\delta\,$, $\frac{1}{2}$$\pm\,$$\delta\,$, 0).
As an example, the magnetic propagation wave vector \emph{q} is
deduced to be ($\frac{1}{2}$$\pm$0.01, $\frac{1}{2}$$\pm$0.01, 0)
for KCrF$_3$ at 47.7 K. However, regarding to the propagation vector
($\frac{1}{2}$$\pm\,$$\delta\,$, $\frac{1}{2}$$\pm\,$$\delta\,$, 0),
there exist several possible magnetic structural schemes including
the sinusoidal and helimagnetic solutions. Polarized neutron
experiments on the KCrF$_3$ single crystals may shed light on the
nature of that incommensurate magnetic structure. Usually, the
presence of modulated phases is the consequence of the competing
next nearest-neighbor exchange interaction and anisotropy in the
spin Hamiltonian \cite{Kocinski, Bak}. In the monoclinic phase of
KCrF$_3$, the tilt of CrF$_6$ octahedra lead to a nonzero angle
between the occupied \emph{e}$_g$ orbital and the (220) plane. As a
result, the Cr-F-Cr superexchange angle deviates from
180$\textordmasculine$ and probably weakens the FM interactions
between Cr$^{2+}$ and four nearest neighbor Cr$^{2+}$ ions inside
the (220) plane. The next nearest-neighbor exchange may play a role
to stabilize the incommensurate AFM state. The paramagnetic state is
established when the sample is heated up to 82 K. The Neel
temperature \emph{T}$_N$ (79.5 K) is clearly defined by the distinct
anomaly in the slope of the C$_P$ data as shown in the inset of Fig.
4b.

In summary, we have investigated the correlation among the
structural transition, orbital ordering and magnetic ordering in
perovskite chromite KCrF$_3$. In this system, two different orbital
ordering states are observed below and above the structural
transition temperature. With decreasing temperature, the
incommensurate \emph{A}-type AFM state emerges firstly due to the
ordering of the moment of Cr$^{2+}$ and the commensurate AFM order
develops from the incommensurate AFM state at 45.8 K with a magnetic
propagation vector change from ($\frac{1}{2}$$\pm\,$$\delta\,$,
$\frac{1}{2}$$\pm\,$$\delta\,$, 0) to ($\frac{1}{2}$, $\frac{1}{2}$,
0). The noncollinear canted magnetic structure and weak
ferromagnetic structure might be the nature of lower temperature
phase zone. The rich magnetic phase diagram with various
commensurate and incommensurate phases is ascribed to competing
nearest-neighbor and next nearest-neighbor exchange interactions,
single ion anisotropy as well as the DM interaction.

\vspace{0.2 cm} The authors are grateful to P. Meuffels and H.
Bierfeld for providing assistance with sample preparation. We also
thank E. Kentzinger and B. Schmitz for the help with the
magnetization and heat capacity measurements.

\appendix


\begin{thebibliography}{10}

\bibitem{Imada}
M. Imada, A. Fujimori, and Y. Tokura, The Physics of manganites:
Structure and transport. Rev. Mod. Phys. \textbf{70}, 1039 (1998).

\bibitem{Lee}
P. A. Lee, N. Nagaosa, and X. Wen, Doping a Mott insulator: Physics
of high-temperature superconductivity. Rev. Mod. Phys. \textbf{78},
17 (2006).

\bibitem{Tokura1}
Y. Tokura and N. Nagaosa, Orbital Physics in Transition-Metal
Oxides. Science, \textbf{288}. 462 (2000).

\bibitem{Tokura2}
Y. Tokura, Rep. Prog. Phys. \textbf{69}. 797 (2006).


\bibitem{Asamitsu}
A. Asamitsu, Y. Moritomo, Y. Tomioka, T. Arima, and Y. Tokura,
Nature(London) \textbf{373}, 407 (1995).

\bibitem{Jin}
S. Jin, M. McCormack, T.H. Tiefel, and R. Ramesh, J. Appl. Phys.
\textbf{76}, 6929 (1994).

\bibitem{Helmolt}
R. von Helmolt, J. Wecker, B. Holzapfel, L. Schultz, and K. Samwer,
Phys. Rev. Lett. \textbf{71}, 2331 (1993).

\bibitem{Wollan}
E. O. Wollan and W. C. Koehler, Phys. Rev. \textbf{100}, 545 (1955).

\bibitem{Margadonna1}
S. Margadonna and G. Karotsis, J. Am. Chem. Soc. \textbf{128}, 16436
(2006).

\bibitem{Rodriguez1}
J. Rodr\'{\i}guez-Carvajal, Physica B \textbf{192}, 55 (1993).

\bibitem{Yoneyama}
S. Yoneyama, K. Hirakawa, J. Phys. Soc. Jpn. \textbf{21}, 183
(1966).


\bibitem{Chatterji}
T. Chatterji, P. F. Henry, and B. Ouladdiaf, Phys. Rev. B
\textbf{77}, 212403 (2008).


\bibitem{Rodriguez2}
J. Rodr\'{\i}guez-Carvajal, M. Hennion, F. Moussa, A. H. Moudden, L.
Pinsard, and A. Revcolevschi, Phys. Rev. B \textbf{57}, R3189
(2008).


\bibitem{Sub}
G. Sub\'{\i}as, J. Herrero-Mart\'{\i}n, J. Garc\'{\i}a, J. Blasco,
C. Mazzoli, K. Hatada, S. Di Matteo, and C. R. Natoli, Phys. Rev. B
\textbf{75}, 235101 (2007).


\bibitem{Scatturin}
V. Scatturin, L. Corliss, N. Elliott, J. Hastings, Acta Crystallogr.
\textbf{14}, 19 (1961).


\bibitem{Giovannetti}
Gianluca Giovannetti, Serena Margadonna, and Jeroen van den Brink,
Phys. Rev. B \textbf{77}, 075113 (2008).

\bibitem{Xu}
Yuanhui Xu, Xianfeng Hao, Minfeng Lv, Zhijian Wu, Defeng Zhou, and
Jian Meng, J. Chem. Phys. \textbf{128}, 164721 (2008).

\bibitem{Solovyev}
Igor Solovyev, Noriaki Hamada, and Kiyoyuki Terakura, Phys. Rev.
Lett. \textbf{76}, 4825 (1996).

\bibitem{Feinberg}
D. Feinberg, P. Germain, M. Grilli, and G. Seibold, Phys. Rev. B
\textbf{57}, R5583 (1998).

\bibitem{Kanamori}
J. Kanamori, J. Phys. Chem. Solids \textbf{10}, 87 (1959).

\bibitem{Anderson}
P. W. Anderson, Phys. Rev. \textbf{109}, 1492 (1958).

\bibitem{Goodenough}
J. B. Goodenough, \emph{Magnetism and the Chemical Bond}
(Interscience Publ., New York, 1963).

\bibitem{Kugel}
K. I. Kugel and D. I. Khomskii, Zh. Eksp. Teor. Fiz. \textbf{64},
1429 (1973) [Sov. Phys. JETP \textbf{37}, 725 (1973)]



\bibitem{Khomskii}
D. I. Khomskii and K. I. Kugel, Phys. Rev. B \textbf{67}, 134401
(2003).



\bibitem{Schweika}
W. Schweika and P. B\"{o}ni, Physica B \textbf{297}, 155 (2001).


\bibitem{Matsumoto}
G. Matsumoto, J. Phys. Soc. Jpn. \textbf{29}, 606 (1970).

\bibitem{Skumryev}
V. Skumryev, F. Ott, J. M. D. Coey, A. Anane, J. P. Renard, L.
Pinsard-Gaudart, and A. Revcolevschi, Eur. Phys. J. B. \textbf{11},
401 (1999).

\bibitem{Talbayev}
Diyar Talbayev, L\'{a}szl\'{o} Mih\'{a}ly, and Jianshi Zhou, Phys.
Rev. Lett. \textbf{93}, 017202 (2004).

\bibitem{Moriya}
T. Moriya, in \emph{Magnetism}, Edited by G. T. Rado and H. Suhl,
(Academic Press, New York, 1984).

\bibitem{Dzyaloshinsky}
I. Dzyaloshinski, J. Phys. Chem. Solids \textbf{4}, 241 (1958).

\bibitem{Kocinski}
J. Kocinski, \emph{Commensurate and Incommensurate Phase
Transitions} (Elsevier, Amsterdam, 1990).

\bibitem{Bak}
P. Bak, Rep. Prog. Phys. \textbf{45}, 587 (1982).


\end{thebibliography}
\end{document}